%% file: 2021_JSS_AIQuality.tex
\documentclass[sigconf]{acmart}

\usepackage{natbib}
\usepackage{graphicx}
\usepackage{array}
\usepackage{multirow}
\usepackage{verbatim}
\usepackage{hyphenat}
\usepackage{tcolorbox}
\usepackage{enumitem}
\usepackage{balance}
\usepackage{todonotes}
\usepackage{threeparttable}
\usepackage{epstopdf}
\usepackage{footmisc}
\usepackage{capt-of}
\usepackage{url}

\AtBeginDocument{%
  \providecommand\BibTeX{{%
    \normalfont B\kern-0.5em{\scshape i\kern-0.25em b}\kern-0.8em\TeX}}}

\setcopyright{acmcopyright}
\copyrightyear{2022}
\acmYear{2022}
\acmDOI{10.1145/1122445.1122456}

\acmConference[CAIN '22]{CAIN '22: 1st Conference on AI Engineering - Software Engineering for AI}{May 21--22, 2022}{Location to be announced}
\acmBooktitle{CAIN '22: 1st Conference on AI Engineering - Software Engineering for AI,
  May 21--22, 2022, Location to be announced}
\acmPrice{15.00}
\acmISBN{978-1-4503-XXXX-X/18/06}

\begin{document}

\title {What is Software Quality for AI Engineers? Towards a Thinning of the Fog}            

\author{Valentina Golendukhina}
\affiliation{%
  \institution{University of Innsbruck}
  \city{Innsbruck}
  \country{Austria}
}
\email{valentina.golendukhina@uibk.ac.at}

\author{Valentina Lenarduzzi}
\affiliation{%
  \institution{University of Oulu}
  \city{Oulu}
  \country{Finland}
}
\email{valentina.lenarduzzi@oulu.fi}

\author{Michael Felderer}
\affiliation{%
  \institution{University of Innsbruck}
  \city{Innsbruck}
  \country{Austria}
}
\email{mfelderer@uibk.ac.at}

\renewcommand{\shortauthors}{Golendukhina V. et al.}

\begin{abstract}
It is often overseen that AI-enabled systems are also software systems and therefore rely on software quality assurance (SQA). Thus, the goal of this study is to investigate the software quality assurance strategies adopted during the development, integration, and maintenance of AI/ML components and code. We conducted semi-structured interviews with representatives of ten Austrian SMEs that develop AI-enabled systems. A qualitative analysis of the interview data identified 12 issues in the development of AI/ML components. Furthermore, we identified when quality issues arise in AI/ML components and how they are detected. The results of this study should guide future work on software quality assurance processes and techniques for AI/ML components.
\end{abstract}




\maketitle

\input{Section/Introduction.tex}
\input{Section/RelatedWork.tex}
\input{Section/EmpiricalStudy.tex}
\input{Section/Results.tex}

\input{Section/Discussion.tex}
\input{Section/Threats.tex}

\input{Section/Conclusion.tex}

\bibliographystyle{ACM-Reference-Format}
\balance
\bibliography{sample}

\end{document}

%% file: Section/Introduction.tex
\section{Introduction}
\label{sec:Introduction}
Artificial Intelligence (AI) is getting more and more popular, being adopted in numerous applications and technologies we use on a daily basis. 
A large number of AI-enabled applications are produced by developers without proper training on software quality practices or processes, and in general, lack state-of-the-art software engineering processes. 

An AI-enabled system is a software-based system that comprises AI/ML components besides traditional software components~\cite{felderer2021quality}. As any software system, AI-enabled systems require attention to software quality assurance (SQA) in general and code quality in particular~\cite{nasa}. 


Current development processes, and in particular agile development models, enable companies to decide the technologies to adopt in their system in a later stage. Therefore, it is hard to anticipate if a system, or if a data pipeline used to develop AI will produce high-quality models~\footnote{http://www.nessi-europe.com/files/NESSI\%20-\%20Software\%20and\%20AI\%20-\%20issue\%201.pdf}. 

One of the reasons for that is due to the fact that the AI engineer is a new profession, and currently there is a very limited number of training or guidelines on issues (such as code quality or testing) for AI and applications using AI code. 
According to preliminary studies~\cite{Masuda2018, LenarduzziAI2021}, developers' training is one of the biggest lacks in software quality assurance for AI, which usually brings several issues related to low quality of AI code as well as low long-term maintenance. Moreover, the software quality of AI-enabled system is often poorly tested and of very low quality~\cite{Wang2020, LenarduzziAI2021}.

The goal of this paper is to investigate SQA strategies adopted during the development, integration, and maintenance of AI components. We conducted an exploratory study, by means of interview, among ten Austrian companies that develop AI-enabled systems.


The insights in this paper enable researchers to understand possible problems on the quality of AI systems, opening new research topics, and allows companies to understand how to better address quality issues in their systems.

The remainder of this paper is structured as follows: Section~\ref{sec:relWork} reports on related works, while Section~\ref{sec:CaseStudy} describes the empirical study design. Section~\ref{sec:Result} presents the results of the conducted interviews. The results are further discussed in Section~\ref{sec:Discussion}. Section~\ref{sec:Threat} identifies the threats to validity, and Section~\ref{sec:Conclusion} draws the conclusion and highlights the future works. 

%% file: Section/RelatedWork.tex
\section{Related Work}
\label{sec:relWork}
The role of Artificial Intelligence in Software engineering has been investigating in the last years, focusing on challenges and best practices~\cite{Amershi2019,nascimento2020software}. However, a limited number of peer-reviewed works highlighted the quality issue in AI-enabled software~\cite{Murphy2006,zhang2020machine,Wang2020}. 
Murphy et al.~\cite{Murphy2006} proposed a testing framework for Machine Learning (ML), introducing the concept of regression testing and an approach for ranking the correctness of new versions of ML algorithms. Besides the proposed model, they also highlighted conflicts in the technical terms with very different meanings to machine learning experts than they do to software engineers (e.g. ``testing'', ``regression'', ``validation'', ``model'').

Nakajima~\cite{Chen15}, in his invited talk, called attention to the product quality of ML-based systems,  identifying new challenges for quality assurance methods. He proposed to identify new testing methods for ML-based systems, such as adopting Metamorphic testing as a pseudo oracle approach and using golden outputs as testing values. 

Several authors conducted empirical studies in which they examined the code of ML projects on bugs or technical debt to discover the difficulties developers face. Sun et al.~\cite{sun2017empirical} analyzed three projects on Github and distinguished seven categories of bugs. According to this classification, the most frequently occurring bugs are connected with compatibility, variables, design defects, and documentation. Additionally, they noticed, that some of the defined bugs are similar to the traditional software projects, and, therefore, can be addressed with techniques adopted in non-ML projects. 

Zhang et al.~\cite{zhang2018empirical} explored bug-related requests on StackOverflow to identify repeated problems in the source code with the focus on TensorFlow-based code and classify these problems. They concluded that the majority of the bugs are related to the incorrect usage of TensorFlow packages and models and resulted in performance inefficiency and functional incorrectness. Nevertheless, 13.6\% of all the bugs were connected with programming issues unrelated to TensorFlow, e.g., Python-related problems. As more frequently faced issues, they distinguished the issues connected with new releases of libraries and problems related to model construction, such as unmatched array size, and incorrect model parameters or structure.  

Islam et al.~\cite{islam2019comprehensive} focused on the bug patterns in different ML libraries, such as Tensorflow, Keras, Theano, Caffe, and Torch, compared their behavior, and identified common antipatterns. In their next study, Islam et al.~\cite{islam2020repairing} further elaborated the classification of programming bugs in DL code and stated that bug patterns of non-ML projects significantly differ from those for ML projects.

Wang et al.~\cite{Wang2020} analyzed Jupyter notebooks, investigating their code quality. They also report that poor coding practices, as well as the lack of quality control, might be propagated into the next generation of developers. Hence, they argue that there is a strong need to identify quality practices, but especially a quality culture in the ML community. Unlike the studies mentioned above that examined the code related issues via code analysis, we looked at the problem of code quality from the developers' perspective to gather their experience and expertise. 

Kim et al. ~\cite{kim2017data} conducted a large-scale study among Microsoft data scientists, identifying the main challenges and best practices. Unlike our study, their study contains a broad view of challenges and possible best practices at every level of interaction with AI systems. Moreover, they describe a perspective of a large multinational corporation, Microsoft, and it can be different from SMEs and startups since the processes in big companies are more mature and developed. 

Another large-scale survey by Ishikawa and Yoshioka ~\cite{ishikawa2019engineers} on challenges for the engineering of ML systems included 278 respondents and examined challenges reported at every stage. They focus on the perception of current strategies and processes in comparison to the traditional ones and challenges that are specifically new due to the nature of ML. We aimed at covering all possible challenges related to product development and inner quality, including challenges that can be found in traditional software. 

Vignesh~\cite{Vignesh} proposed to continuously validate the quality of ML models, considering black box techniques and evaluating the performance of model post-deployment on test data sets and new data from production scenarios. He also proposes to adopt metamorphic testing, involving data mutation to augment test data sets used for evaluating model performance and accuracy. 

Lwakatare et al.~\cite{Lwakatare2019} performed the work closest to this work. They discussed software engineering challenges based on a survey conducted on six companies. The result is a taxonomy of challenges that  consider the life-cycle activities of ML code (such as assemble data set, create model, (re)train and evaluate, deploy). However, differently than in our work, they did not identify clear issues and possible solutions. 

Lenarduzzi et al.~\cite{LenarduzziAI2021} investigated the main software quality issues of Artificial Intelligence systems and drafted a quality taxonomy for AI-based systems. Results revealed that the developers' training is one of the biggest lacks in AI, which usually implies several issues related to low code quality of AI-code as well as low long-term maintenance.  

There is research done on the topic of software engineering solutions for AI-based components. Santhanam~\cite{santhanam2020quality} provided a list of traditional activities modified to AI development purposes among which there are manual inspections, static analysis, white and black box testing, data assessment, monitoring, and debugging techniques. Breck et al.~\cite{breck2017ml} collected 28 testing and monitoring strategies in use for features and data, models, infrastructure at every stage of development. The described techniques are adapted for ML components and represent a checklist of steps to improve the quality using the tools already developed and applied in traditional SE. Serban et al.~\cite{serban2020adoption} elaborated SE practices of Santham et al.~\cite{santhanam2020quality} further and presented 29 engineering best practices for ML. 

%% file: Section/EmpiricalStudy.tex
\section{The Exploratory Study}
\label{sec:CaseStudy}

\subsection{Goal and Research Questions}
The \emph{goal} of this study is  to investigate the software quality assurance (SQA) strategies adopted during the development, integration, and maintenance of AI/ML components. 

Based on our goal, we defined the following two research questions (RQs): 

\vspace{2mm}
\begin{tabular}{p{0.5cm}p{7cm}}
\textbf{RQ$_1$} & Which internal quality issues have been experienced by industry in AI-enabled systems? \\
\textbf{RQ$_2$} &  Which quality practices are applied in the industry for keeping the quality of the AI/ML code under control? \\
\end{tabular}

\begin{table*}[t]
\centering
\small
\caption{The Survey}
\label{tab:Survey}
\begin{tabular}{p{0.5cm}|p{15cm}} \hline 
\textbf{RQ} & \textbf{Questions} \\ \hline 
\multicolumn{2}{c}{\textbf{Profiling}} \\ \hline 
& Your Company \\
& Your role \\
& Size of your team (number of developers and operators) \\
& Which type of software do you develop?\\
& Which is the role of AI in your software?\\
& Which is the development process for AI components adopted in your company?\\
& Could you provide some metrics concerning the development process of your AI components? (number of commits, days per sprint in case of agile, number of pull requests, number of developers per issue, etc.)\\ \hline 
\multicolumn{2}{c}{\textbf{SQA strategies}}\\ \hline
RQ$_1$ & Which issues have you experienced in your AI/ML code? \\
RQ$_1$ & Which quality issues do you commonly face in AI/ML code? \\
RQ$_1$ & How do you discover quality issues in AI/ML code during the development process?\\
RQ$_1$ & At which phase you mainly discovered quality issues (development, integration, or maintenance)?\\
RQ$_1$ & At which development/integration phase(s) do you evaluate the AI/ML component code quality?\\ \hline 
RQ$_2$ & Which quality practices do you apply in your system? \\
RQ$_2$& Which quality practices do you apply for your AI/ML code? \\
RQ$_2$& Do you adopt SQA standards? Which?\\
RQ$_2$& How do you test your AI/ML component?\\
RQ$_2$& How do you test the integration of your AI/ML component into the system?\\
RQ$_2$& How do you select the AI/ML component to integrate in your system? Is it based on a code quality evaluation?\\ \hline 
\end{tabular}
\end{table*}

\vspace{2mm}
In RQ$_1$, we aim at characterizing the main issues faced by companies developing AI-enabled systems. We are considering companies embedding AI-enabled libraries or developing software that makes powered by ML. 
In RQ$_2$, we want to understand which practices have been applied to solve the issues faced in RQ$_1$. As an example, companies could apply well-known QA practices, or different types of workarounds.

\subsection{Study Design}
Our empirical study is an exploratory survey by means of semi-structured interview with representatives of ten Austrian SMEs that develop AI-enabled systems. 

At the beginning of each interview, we briefly introduced the project and the goals of the study to an interviewee. 
The first part of the interviews comprises general questions to shape a profile of the respondent’s company, his or her role in the company, the company’s domain, and the role of AI in it. Then, we introduced the questions on the main topic including the questions about the development process, main challenges, quality issues, ways to discover quality issues, and implemented quality practices. 

\subsubsection{Interview Design} The interview  is composed by the following two main sections according to our RQs. The interview  including all the questions is reported in Table~\ref{tab:Survey} and in the replication package~\footref{package}.

\begin{itemize}
\item \textit{\textbf{Demographic information}}: We collected demographic background information such as role and relative experience. We also collected company information such as application domain, organization’s size via number of employees, and number of employees in the respondents' own team.
\item \textit{\textbf{SQA strategies}}: We collected information about which issues do the developers experience in their programming code for AI/ML components, and what do they do to prevent and solve such issues. In particular, we were interested in SQA techniques the developers adopt in their development process in every stage of the developmental process i.e., development, integration, and maintenance.
\end{itemize}  

\subsubsection{Participants Selection} We considered the Austrian AI Landscape\footnote{\label{AustrianLanscape}https://www.enlite.ai/insights/ai-landscape-austria}: 250 Austrian companies and research institutions dealing with AI presented by enliteAI. Austrian AI Landscape is characterized by steady growth in recent years with a 20\% average annual growth in the last 6 years. This market area mostly populated with startups can provide great insights into AI and ML-based components development. 
The companies we were interested in had to either have AI as a core component of their activity or develop AI-enabled components to support the main non-AI-related activity. Ten companies have agreed to an in-depth interview. 
Out of 250 organizations in the list, we tried to reach 95 companies that were suitable for the purposes of our research. 
We did not consider research institutions, universities, conferences, business incubators, or governmental funds in order to focus on industry practitioners. Moreover, before the participation, the companies were informed of the study goal. 

\subsection{Study Execution} 
We conducted our interview  from March to April 2021. Due to the COVID-19 pandemic, the one-to-one interviews took place online via Google Meets, Microsoft Teams, and Zoom. Each interview was conducted by one researcher and lasted from 30 to 50 minutes. The interviews were recorded and then transcribed for qualitative analysis. 
Since the interviews could be held online, we could reach companies all over Austria, resulting in four companies from Vienna, two companies with sites in Tyrol and Upper Austria, and two companies from Styria.
To better understand the topic, we conducted the interview with both employees working directly on the AI code and heads of departments to gain insight into the topic in a broader sense. Nevertheless, it was crucial for us that all of the participants are closely involved in the development process of AI/ML components. 

\subsection{Data Analysis}
One author manually transcribed the record and two authors inspected the answers of each interview and then provided a hierarchical set of codes from all the transcribed answers, applying the open coding methodology~\cite{Wuetherick2010}. The authors discussed and resolved coding discrepancies and then applied the axial coding methodology~\cite{Wuetherick2010}. 
Open questions were analyzed via open and selective coding~\cite{Wuetherick2010}. The answers were interpreted by extracting concrete sets of similar answers and grouping them based on their perceived similarity.

\subsection{Verifiability and Replicability}
We made available in our online appendix\footnote{\label{package} \url{https://figshare.com/s/6c26e7d6e90c1d464c44}} the complete raw data, including the interviews. 

%% file: Section/Results.tex
\section{Results}
\label{sec:Result}
In this Section, we report the obtained results in order to answer to the Research Questions. 
First, we depicted the respondents' background and the development process adopted in the interviewed companies.  

\vspace{2mm}
\textbf{Background.} We interviewed ten companies affiliated to the Austrian AI Landscape\footref{AustrianLanscape} (Table~\ref{tab:Company}). The vast majority (90\%) of the respondents report that data science and software engineering departments are separated in their companies. However, in most of the cases (50\%), the team dealing with AI consisted of less than 10 employees. 
All companies represented SMEs, and the number of employees ranged from 8 to 35 employees from different organizational domains such as robotics, health, or finance.

\begin{table*}[t]
\centering
    \caption{Companies description}
    \label{tab:Company}
 \small
\begin{tabular}{l|l|r|l} \hline 
\textbf{Id} & \textbf{Domain} & \textbf{Org. Size} & \textbf{Interviewee Role} \\ \hline 
C1 & Industry \& Robotics & $\approx$ 10 & CTO \\
C2 & Health \& MedTech & $\approx$ 20 & Leading ML Developer \\
C3 & Language \& Text & $\approx$ 35 & Research Engineer \\
C4 & Language \& Text & $\approx$ 10 & ML Team Leader \\ 
C5 & Industry \& Robotics & $\approx$ 35 & CEO, Head of Development \\
C6 & ERP & $\approx$ 4 & Data Science Team Leader \\
C7 & Health \& MedTech & $\approx$ 32 & Data Science Team Leader \\
C8 &  Platform & $\approx$ 20 & Head of Development \\ 
C9 & Industry \& Robotics & $\approx$ 30 & Data Scientist \\
C10 & FinTech & $\approx$ 27 & CEO \\ \hline 
    \end{tabular}
\end{table*}



The role of AI varies across different companies. It is used for defect-recognition in production, natural language processing, route planning, business planning such as resource allocation or customer’s purchases prediction, quality maintenance, disease prediction in healthcare, and real estate evaluation. 9 out of 10 companies stated that AI plays the core role in their company’s activity and there is not a single solution without an AI/ML component. One company was developing a platform for AI/ML model development, support, testing, and usage. 

\vspace{2mm}
\textbf{Development process}. To generally understand the context and the role of AI/ML in the organization of the work of our respondents, we asked them to talk about the development process within the company and how AI/ML was embedded into it. Asking about the development process within the company, we aimed to get some information about both technical details of the development such as programming language, AI/ML frameworks, software or other tools the respondents found important to share with us, and organization of the development process, e.g., workflow, project management frameworks, or working strategies. 

All respondents told us that they use agile or agile-like software engineering processes with one to two-week sprints. However, 60\% communicated some planning-related issues associated with the impossibility of identifying hard deadlines in AI/ML development, especially in the early stages of development. Several respondents noted that it is important to invest the right amount of time from the beginning to not have negative consequences in the future. 





Since the research is a large part of AI components development, it is vital to adapt the organizational workflow in its favor. One way reported was to shorten the sprints at the beginning of the development process to have more flexibility with changes. Another way to overcome the planning-related issues is to introduce spikes in the development processes for time-ambiguous activities, including research in AI/ML.

In the second part of the question, we also asked the participants to tell us about the tools used in the company for data science tasks. Python is the most common programming language, which was reported by 80\% of the companies. However, several companies reported more than one language including R, Julia, and MATLAB. Several companies also mentioned deep learning frameworks such as PyTorch, TensorFlow, and Onnx. However, companies often try not to rely too much on the external libraries or frameworks, which we will discuss more in detail in the next subsection. 

\subsection{Issues in AI/ML Code (RQ$_1$)}
\begin{figure}
    \centering
    \includegraphics[width=1.0\linewidth]{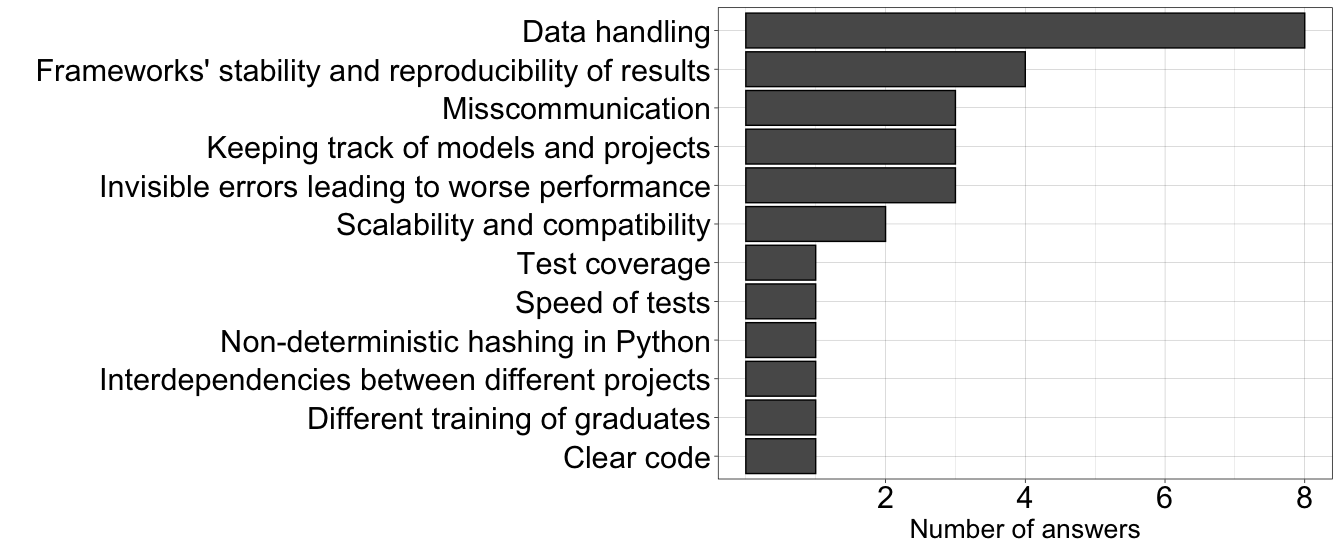}
    \caption{Issues in development of AI-enabled systems}
    \label{fig:Issue}
\end{figure}

In this question, we asked the participants about the problems they experience in their AI/ML code and common quality issues. Although we clarified that the main interest of our study is AI/ML code, it was difficult for the respondents to distinguish between the quality of the code and the quality of data and models in the case of AI/ML components. In total, we collected 12 different categories of issues the respondents experienced in their AI/ML code. 
The issues and their frequency are represented in Figure~\ref{fig:Issue}.

The most frequently recalled issue is data handling. It was named as the biggest issue in AI/ML code development by 8 companies. Since data quality is out of the scope of this study, we did not go deeper into this topic during the interviews and classified all the different data quality issues into the category of data handling. It includes poor labeling, poor dataset composition, data management, data preparation, and pipeline.

\vspace{2mm}
\begin{centering}
\textit{``A big part of all AI components is the data preparation and the pipeline. If you have poor quality data, all these preparation steps might be a huge problem. 80\% of the effort is data preparation and cleaning. With our platform, we can provide more standardized processes to have cleaned data for the training process.'' (C8)}

\vspace{2mm}
\textit{``Problems with the data handling is our job’s description.'' (C7)}

\end{centering}
\vspace{2mm}

The second issue which was named by 4 companies is the stability of the frameworks and the results they reproduce. The practitioners talked about fast changes in machine learning frameworks that influence the overall performance of the system. Therefore, all of the respondents who named it a problem also confirmed that they do not rely on external libraries too much. For some companies, it is connected with security issues and every library they use has to be checked by the IT security department before implementing it in a project. For other companies, keeping one version of a package was the solution. 

\vspace{2mm}
\begin{centering}

\textit{``We are trying to be conservative in relying on external libraries for the reason that they are not stable. We just freeze the package version and stick with one tool or library that we know works and try to minimize dependencies on things. We only use things that are very well tested and widely used.'' (C7)}

\end{centering}
\vspace{2mm}

The third most frequently mentioned issue was miscommunication. By miscommunication, the participants meant the inability of data scientists to communicate  effectively with software engineers or with the clients or business experts. Because data science is separated from the software engineering discipline, although it also requires to engineering software, data scientists have their specific vocabulary often not clear for the experts from the software engineering domain and vice versa, confirming the misleading terminology between software engineering and AI/ML~\cite{LenarduzziAI2021}. For instance, the term "regression" typically refers to a statistical regression model in data science, but to regression testing in software engineering. The problem of miscommunication is encountered even among data scientists with different educational backgrounds, e.g., math and economics~\cite{LenarduzziAI2021}. 

Two respondents shared the experience of solving this problem within their firm. In both cases, the specialists from the communicating departments, such as data science and medical experts collaborated to create a vocabulary of frequently used or misunderstood terms to increase the efficiency of their communication. 

\vspace{2mm}
\begin{centering}

\textit{``We spent a ridiculous amount of time defining it that everyone agrees on it, but it was worth it. Now we are on the same page.'' (C7)}

\end{centering}
\vspace{2mm}

Invisible errors or errors leading to decreased performance were also indicated by three companies. Several participants noted that although programming bugs also take place, they are usually easy to identify, in contrast to hidden errors connected with models and uncertainty. The latter are difficult to identify and hence debug. One of the respondents added that such bugs can stay undetected for several months. They do not lead to failure, but less accurate results. 

Keeping track of models and projects was also identified by three respondents and relates to the decision-making process of choosing the right model or being able to reuse a model developed in the past. The scalability of models describes the issues two companies had in their experience. These are the scalability of a model used by many users simultaneously and the scalability of a model on growing GPU or CPU power.
The remaining six issues appeared one time each during all the interviews. Two of them are related to testing. These are test coverage and speed of tests, where the latter relates to the model tests. We will closely discuss testing strategies in the following parts. 

One participant raised the issues of different training of engineering and economics graduates and clear code, and these can be connected. He implied that graduates with an engineering degree are more familiar with best practices and programming techniques. Therefore, they can better organize their code than economics graduates. He highlighted the importance of “a clear, maintainable, modular and well-documented code” in production, which cannot be easily achieved by specialists without proper training. Moreover, it is challenging to achieve while whiting the code via APIs:

\vspace{2mm}
\begin{centering}

\textit{``We are solving a lot of problems via APIs. We use frameworks, and it is like a drug: you start with a framework like Shiny, and then it is really hard to come to a clear code because it makes it so easy to produce a lot of code.'' (C10)}

\end{centering}
\vspace{2mm}

The last two issues describe the code-related problems. Those are interdependencies between different projects, which means that the change in one part of the source code may influence the other parts without a person being acknowledged of it. Therefore, highlighting the interdependencies between different products is necessary. Non-deterministic hashing in Python is a tool-specific issue that leads to the change of items in some of the instances.

Since we were conducting an explorative study, we tried to influence the participants as little as possible. We navigated them through the questions, but let them speak about the issues they found to be important. Data and models’ quality is undoubtedly an important issue in the development of AI/ML components. It is tightly connected with code quality, and one can hardly be investigated without another. This might be the reason why, although we wanted to investigate code quality, most of the answers were tightly connected with the quality of data and models. 

We tried to specify the difference between code and model-related issues in the next question by asking the practitioners about the common factors that influence the quality of AI/ML components. Then, we grouped the replies into two categories: data- or model-related and purely code-related issues.

As a result, seven participants stated that they do not have major issues connected directly with the code. Model and data have the biggest influence on the quality, whereas bugs in AI/ML code can be easily detected and fixed and therefore are not perceived as quality threads. 

\vspace{2mm}
\begin{centering}

\textit{``That is not related to AI code, we have code quality issues, but we also have automated testing so that’s not an issue in the end because it is more or less fixed and covered.'' (C5)}

\vspace{2mm}
\textit{``I did not have that many problems with the code. Most of the time, when I am reading through the project, it is easy to read and understand. I mostly have problems with the data.'' (C9)}

\end{centering}
\vspace{2mm}

The other three respondents stated that they might experience generic programming quality issues. The quality and readability of the code might suffer, they also experience basic problems with variables and not sufficient documentation. Furthermore, error handling is sometimes not done well. 

\vspace{2mm}
\begin{centering}

\textit{``Error handling is not done well. People do not think like engineers, and they do not care if the service has to be alive.'' (C10)}

\end{centering}
\vspace{2mm}

The majority of the respondents stated that the main challenge in AI/ML components’ code is due to data and model-related issues. Nevertheless, the other respondents say that the code itself can be an issue. 

\subsection{Quality Practices for AI/ML Code (RQ$_2$)} 
Answering to this RQ, we drive the discussion, focusing on when quality issues arise in AI/ML components and how they are detected. 

\vspace{2mm}
\textbf{When}. We investigated when do the participants face the most quality issues and when do they evaluate the quality of the AI/ML component. The number of answers exceeds 10 because the respondents could choose more than one answer. Most of the respondents (70\%) stated that they experience the most quality issues in the development phase. 40\% and 50\% reported the vast number of issues in integration and maintenance phases, respectively. By quality issues, they meant overall malfunction in the system that leads or could potentially lead to a decreased performance. For instance, an issue in the maintenance stage mentioned by several respondents was the inconsistency of the data used for testing and in practice. 

Based on the previous answer, we asked at which phase they check the quality of the code for AI/ML components. The answers did not contradict the previous question. Most of the evaluation process for AI/ML-based code is done in the development phase, i.e., during the development of the algorithms or the learning phase. It was stated by all the participants. 

Some participants confirmed that they evaluate the quality of AI/ML components in the integration phase either. In this case, it was about checking the overall system performance, including special checks for the AI components as a part of it. 

Yet several participants noted that they introduced continuous integration in their organizations, and it makes it different to tell the clear difference between the development phases, i.e., development and integration.

\begin{figure}
    \centering
    \includegraphics[width=1.0\linewidth]{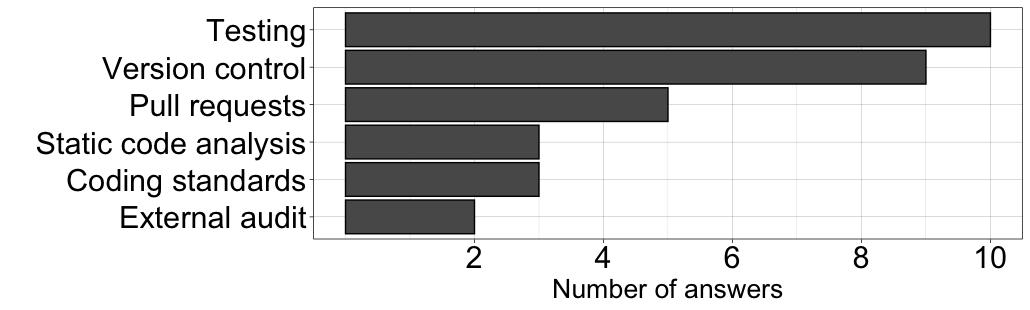}
    \caption{Quality practices applied to AI-enabled system}
    \label{fig:How}
\end{figure}

\vspace{2mm}
\textbf{How.} We found out at which point the quality issues are being detected, we can focus on how they can be discovered during the development and integration processes (Figure~\ref{fig:How}). To investigate the practices the companies apply to AI/ML code, we have decided first to analyze the overall quality practices applied to the system and then how they are further applied to the AI/ML code. We asked our respondents to name the main quality practices they use in their systems and got 6 main categories. 

Testing was mentioned by every respondent, and version control by nearly every respondent (90\%). Pull requests were indicated by 50\% of the participants. Static code checking and using coding standards such as ISO or internally developed code standards were reported by three participants each. The external audit was an obligatory part of software quality assurance for two companies because of special regulations in the domain for which they create their software, such as MedTech.

\begin{figure}
    \centering
    \includegraphics[width=1.0\linewidth]{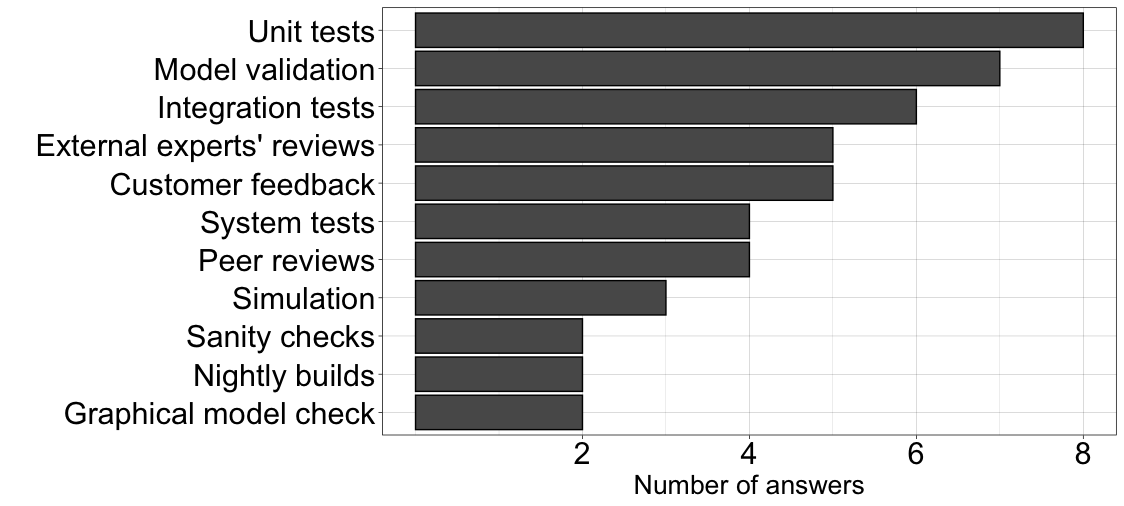}
    \caption{Quality practices for AI components}
    \label{fig:How2}
\end{figure}

Then we analyzed all the techniques and practices the participants use to detect quality issues in AI/ML code. Figure \ref{fig:How2} shows the final 10 categories mentioned at least by two participants. 

Unit tests were the most frequently mentioned practice for quality issues detection in AI/ML code. All the participants pointed out that they use unit tests to identify quality issues, and several respondents mentioned that they use unit tests in the context of test-driven development. 
Although unit tests appeared to be the most common quality practice, some participants raised awareness about the suitability of such tests to the AI/ML development. They stated that because of the highly ambiguous and rapidly changing nature of AI/ML development, unit tests and test-driven development can sometimes be overwhelming and less effective. 

\vspace{2mm}
\begin{centering}

\textit{``I am personally not 100\% sure if unit testing in the machine learning field is very useful. A lot of data science guys follow one idea one week and the next idea next week. Unit tests which they wrote in week number one are completely useless for week number two because the approach is completely different.'' (C9)}

\end{centering}
\vspace{2mm}

The second common answer was model checks including benchmarking with the historical data, checking against several databases, average guesses, experts’ conclusions, etc. Seven participants mentioned it in the interview. In some reported cases, it can be considered as part of unit testing, especially when unit testing is done to check model quality. 

The next category is integration tests, and they were mentioned by 6 participants. Integration tests are necessary to check the work of algorithms on different platforms or environments. Some respondents mentioned that such tests take place within a continuous integration pipeline and are as much automated as possible. All the participants who mentioned that they have to run some tests manually are also working to automate tests as much as possible. 

\vspace{2mm}
\begin{centering}

\textit{``If something has to be tested manually, at least 4 weeks later or two sprints later we have an automated solution. 95\% of tests are automated.'' (C5)}

\end{centering}
\vspace{2mm}

Review by external experts and customers’ feedback both were mentioned by five respondents. By external experts' review, they referred either to their colleagues from another department such as healthcare experts within the firm dealing with AI in the medical domain or to R\&D department feedback after deploying and monitoring models for some time. Also, in several cases, there is a lot of communication and feedback between different teams within the firm, such as data science or software engineering. It was also noted that such feedback can have formal and informal nature. One respondent described this as an ever-changing process, with consultations in person and less formal at the beginning of a project, while at later stages they are organized in the form of formal meetings. In another company, external reviews were organized on a regular basis and each developer had to present and explain her or his code to an external expert at least once a year.

Customers' feedback also plays an important role in quality issues detection, since many research companies work closely with the customers and are able to get continuous feedback from them. Customers can evaluate the quality of a final product, but also be more involved and test every new feature introduced to a product. 
40\% of respondents named peer reviews as a quality check instrument. While one respondent noted that they would like to increase the use of this practice in the organization, another respondent confirmed that they use peer reviews but are trying to reduce it to the necessary minimum.

\vspace{2mm}
\begin{centering}

\textit{``We increasingly do the code reviews, including the senior people checking the code.'' (C10)}

\vspace{2mm}
\textit{``We are moving away from pull requests because we do not have time to do reviews like that on everything. We do it on things that are confusing or complex or uncertain enough that it is worth having a second opinion.'' (C7)}

\end{centering}
\vspace{2mm}

However, every respondent who mentioned peer reviews highlighted their usefulness, especially in difficult or critical cases. 
Testing in simulation is another possible testing strategy that is suitable in some specific domains that deal with pieces of machinery. We will have a closer look at this in the component testing part.

The last three categories were each brought up by two participants. These are sanity checks, graphical model checking, and nightly builds. Sanity checks as well as nightly builds are typical for checking a continuously integrated system. Their application to the code of AI/ML components is logical because they check the functionality of every new integrated element in the system. 

Graphical model checking is an extension of performed model checks with the purpose of graphical representation of model accuracy or other essential metrics for easier control over the system, implemented changes, and existing models. 

While comparing system quality practices adopted in the organization and the techniques to check the quality of AI/ML code, we could notice that the same respondents named the same techniques for both system and AI/ML code quality assurance. Testing remains the most commonly used technique for the system and AI component quality. If the company relies on peer reviews for ensuring the quality of the system, it also used this technique in AI/ML code components analysis. Due to the domain specific nature of data science models, external reviews might play a larger role in the development of AI components and performed more frequently than external reviews of the entire system. Version control and coding standards were only mentioned in the context of system quality and not AI component quality.


%% file: Section/Discussion.tex
\section{Discussion}
\label{sec:Discussion}

SQA for AI-based components is a topic gaining popularity and attracting more and more interest in the last years. The purpose of this study was to, first, understand the challenges the developers often face in the development of AI-based products, and second, examine how SQA methods are adopted by the industry based on the experience of data scientists and software developers from Austrian SMEs and startups. 

As many studies stated before \cite{hutchinson2021towards, islam2019comprehensive, kim2017data}, challenges related to data, e.g., data quality, availability, or processing issues, occur the most frequently and are very time-consuming in the development process of AI-systems. According to our findings, data handling is considered to be the most challenging activity in AI-enabled systems. Thus, much effort is devoted to solving data and model-related challenges and maintaining their quality. Consequentially, testing for ML systems needs more research and development, as there are still many issues that do not have a universal solution. Testing of ML systems requires testing of data and models \cite{arpteg2018software}. However, current testing strategies often cannot be applied to existing problems due to the behavior uncertainty, lack of oracle, and other challenges connected with probabilistic nature of AI components. We found that "invisible errors" that cannot be identified by traditional or already existing testing techniques are a big challenge that affects the final results. Such bugs do not lead to an error but reduce the effectiveness or efficiency of an AI-based component. Zhang et al. closely examined such bugs related to TensorFlow and identified several reasons behind them, such as incorrect model parameters, confusion with computation model, TensorFlow API change, and other \cite{zhang2018empirical}.

Our findings on the bugs in the code confirmed already existing results of studies on bugs. As it was stated by the participant of our study, stability of frameworks is a big challenge. The bugs related to this issue appeared in 17.6\% checked projects on GitHub \cite{islam2020repairing}, which showed the highest rate among other bugs examined on GitHub. Another problem raised in the interviews was compatibility and scalability. The study of Sun et al.  \cite{sun2017empirical} showed that compatibility bugs take place in 22.5\% cases of bugs on GitHub, along with variable and design defect bugs. These challenges can be addressed by existing testing and debugging tools \cite{breck2017ml, serban2020adoption}. Nonetheless, code quality issues take place in challenges, and developers without SE experience might be especially affected by it. Since code in AI-enabled applications has a specific form, most of the developers in our study did not report significant problems in the programming code itself.  This issue can be further examined in relation to the background and experience of respondents. Moreover, coding styling guidelines seem to be one of the solutions for code issues, but they must be further adjusted to AI code. 

After comparing the solutions the developers applied to ensure the quality of the system and AI-components, coding and quality standards were applied only for system quality. Overall, we found that there is a low acceptance of SQA standards in the practice, and testing is the main quality assurance technique. One of the reasons for this can be the lack of standards specifically designed for AI-based products. The existing quality standards issued by ISO are originally aimed at the quality of traditional software products. Moreover, Siebert et al. highlighted the problem of high abstraction of existing SQA standards, which makes them hard to apply in practice and results in lower acceptance by developers \cite{siebert2020towards}. SQA standards covering the quality of traditional development are often unsuitable for the development of AI-enabled systems, or represent abstract categories that cannot be easily implemented in practice. This again highlights the gap between SQA for traditional SE and SE of AI systems. Whereas traditional SE is a mature field of study with a lot of research on the topic of SQA, SE for AI is a relatively new area requiring new approaches and strategies for SQA due to its different nature from traditional software. Already existing tools can be adapted to maintain the quality of AI systems, but they are not sufficient to cover all the aspects necessary to ensure the quality of AI-based products. Acknowledging this gap contributes to the further research in the area that is needed to develop new solutions and achieve certain levels of maturity for SQA practices in AI development.

%% file: Section/Threats.tex
\section{Threats to Validity}
\label{sec:Threat}
In this Section, we discuss the threats to validity, including internal, external, construct, and conclusion validity. We also explain the different adopted tactics~\cite{YinCaseStudies2009}. 

\smallskip
\noindent \textbf{Construct Validity}. The empirical study and the survey design, the execution, and the analysis followed a strict protocol, which allows replication of the survey. However, the open questions were analyzed qualitatively, which is always subjective to some extent, but the resulting codes were documented. 
Participants may act differently than they do during the empirical study~\cite{Wohlin2012}. To mitigate these threats, we clearly explained the purpose of the study and asked the participants to answer preliminary questions that allow us to profile their experience.
The questions are aligned with standard terminology and cover the most relevant characteristics and metrics. In addition, the survey was conducted in interviews, which allowed both the interviewees and the interviewer to ask questions if something was unclear.

\smallskip
\noindent \textbf{Internal Validity}.
A non-rigorous experiment design can affect negatively the results~\cite{Wohlin2012}. To deal with this threat, we design and conducted the empirical study according to the guidelines. We formulated the questions in a way that we have only direct questions requiring as little interpretation as possible, to avoid a misunderstanding that would lead to meaningless answers.  We performed a validation process for the designed survey to detect any inconsistency or misunderstanding before the execution.
One limitation that is always a part of survey research is that surveys can only reveal the perceptions of the respondents, which might not fully represent reality. However, our analysis was performed by means of semi-structured interviews, which gave the interviewers the possibility to request additional information regarding unclear or imprecise statements by the respondents. Although our results cannot reflect the real frequency of the observations, e.g., some challenges mentioned by few respondents might have a higher importance in practice, it proves the fact of the existence of certain phenomena and provides directions for further research. 

\smallskip
\noindent \textbf{External Validity}.
The responses were analyzed and quality-checked by a team of four researchers.
We interviewed ten companies located in Austria. We considered only companies that develop AI-based components. For each company, we selected a profile who has a broad overview of the entire AI development process. The results cannot be generalizable due to the small sample size of participants. Moreover, research in large companies and other countries can deliver different outcomes. Nonetheless, similarities with the findings of other researchers can partly validate the correctness of the findings.

\smallskip
\noindent \textbf{Conclusion Validity}.
The survey design, its execution, and the analysis followed a strict protocol, which allows replication of the survey. However, the open questions were analyzed qualitatively, which is always subjective to some extent, but the resulting codes were documented.

%% file: Section/Conclusion.tex
\section{Conclusion}
\label{sec:Conclusion}
In this paper, we presented an exploratory study to investigate SQA strategies adopted during the development, integration, and maintenance of AI/ML components. For that purpose, we conducted interviews with representatives from ten Austrian SMEs that develop AI-enabled systems. A qualitative analysis of the interview data showed 12 issues in the development of AI/ML components. Furthermore, we identified when quality issues arise in AI/ML components and how they are being detected. 

Future work is to be done to collect more data on the topic and involve more respondents. Moreover, the described challenges and solutions can be further studied and refined to the best practices and generalized solutions. More attention is required to the issues related to the code quality, considering different libraries and programming languages. 

SQA for AI significantly differs from SQA for traditional SE and this area of research is still underdeveloped, although the first steps are already done. The creation and adoption of SE practices to AI development is an important step towards increasing and maintaining the quality of AI systems. Overall, further research in the area and training of the developers can improve the quality dramatically and contribute to future generations of data scientists. The results of this study should guide trainings on SQA of AI/ML developers and future work on software quality assurance processes and techniques for AI/ML components. 

\section*{Acknowledgement}
This work was partially supported by the Austrian Science Fund (FWF): I 4701-N.